\documentclass[seceq]{ptptex}

\usepackage{graphicx}
\usepackage{wrapft}

\newcommand{\gsim}{>\kern-12pt\lower5pt\hbox{$\displaystyle\sim$}}
\newcommand{\lsim}{<\kern-12pt\lower5pt\hbox{$\displaystyle\sim$}}

\newcommand{\beq}{\begin{equation}}
\newcommand{\eeq}{\end{equation}}
\newcommand{\beqa}{\begin{eqnarray}}
\newcommand{\eeqa}{\end{eqnarray}}

\def\sla{\slash{\!\!\!} }
\newcommand{\vp}{\mbox{\boldmath $p$}}

\notypesetlogo  

\title{
Ferromagnetism and Superconductivity in Quark Matter\\
--- Color magnetic superconductivity ---
}

\author{
Toshitaka {\sc Tatsumi}$^{1,}$
\footnote{E-mail: tatsumi@ruby.scphys.kyoto-u.ac.jp} 
Tomoyuki {\sc Maruyama}$^{2,}$
\footnote{E-mail: tomo@brs.nihon-u.ac.jp} and 
Eiji {\sc Nakano}$^{3,}$\footnote{E-mail: enakano@comp.metro-u.ac.jp} 
}

\inst{
$^{1}$Department of Physics, Kyoto University, Kyoto 606-8502,Japan \\
$^{2}$College of Bioresouce Sciences, Nihon University, 
         Fujisawa, 252-8510, Japan\\
$^{3}$ Department of Physics, Tokyo Metropolitan University, 
         1-1 Minami-Ohsawa, Hachioji, Tokyo 192-0397, Japan 
}

\recdate{
}

\abst{
A coexistent phase of spin polarization and color superconductivity 
in high-density QCD is studied at zero temperature.
The axial-vector self-energy stemming from the Fock exchange term 
of the one-gluon-exchange interaction
has a central role to cause spin polarization.
As a significant feature,  
the Fermi surface is deformed by the axial-vector self-energy 
and then rotation symmetry is spontaneously broken down. 
The gap function results in being anisotropic in the momentum space 
in accordance with the deformation.  
It is found that spin polarization barely conflicts with color 
superconductivity, 
and almost coexists with it.
}

\begin{document}

\maketitle

\section{Introduction}
Recently much interest is given for high-density QCD, especially for 
quark Cooper-pair condensation phenomena at high-density quark matter
(called as color superconductivity (CSC)), 
in connection with, e.g., 
physics of heavy ion collisions and neutron stars \cite{CSC1}.
Its mechanism is similar to the BCS theory for the 
electron-phonon system and the quark-quark interaction is 
mediated by colored gluons in CSC; the color anti-symmetric $\bar 3$
channel is the most attractive one.

In this talk we would like to address another aspect expected in 
quark matter: spin polarization or ferromagnetism (FM) of quark matter.
A possibility of ferromagnetism in quark matter 
has been first suggested by one of the authors (T.T.) 
by the use of the OGE interaction  \cite{Tatsu};  
a trade-off between the kinetic and 
the Fock exchange energies gives rise to spin polarization, 
similar to Bloch's idea for itinerant electrons \cite{blo, yos}. 
Salient features of spin polarization in the relativistic system 
are also discussed in Ref.~2).

If these phases are realized in quark matter, there should be some 
interplay between them; 
we examine here a possibility of the coexistence of FM 
and CSC in quark matter.
It would be worth mentioning in this context that  
ferromagnetism (or spin polarization) and superconductivity are 
fundamental concepts in condensed matter physics, and  
their coexistent phase has been discussed for a long time \cite{MagSup1}.
As a recent progress, a superconducting phase have been discovered 
in ferromagnetic materials and many efforts have been made to understand  
the coexisting mechanism \cite{MagSup2}.

In a phenomenological context their coexistence should have some
implications 
for the magnetic and thermal properties of compact stars.
Recently, a new type of neutron stars, called  ``magnetars'', 
with a super strong magnetic field of $\sim O(10^{15}$G)  
has been discovered \cite{MAG2}.
They may remind us of a long-standing problem about the origin of the magnetic
field in compact stars,  
since its strength is too large to regard it 
as a successor from progenitor main-sequence stars, 
unlike canonical neutron stars \cite{MAG3}.
Since hadronic matter develops over inside neutron stars 
beyond the nuclear density ($\rho_0$$\sim$$0.16$${\rm fm^{-3}}$), 
it should be interesting to consider the microscopic origin of the magnetic 
field in magnetars.

Thus, 
it might be also interesting to examine the possibility of the 
spin-polarized phase
with CSC in quark matter, in connection with magnetars.

We investigate spin polarization in the color superconducting phase 
by a self-consistent framework, 
in which quark Cooper pairs are formed 
under the axial-vector mean-field.
We shall see that this phenomenon is a manifestation of 
spontaneous breaking of both color $SU(3)$ and rotation symmetries.

\section{Ferromagnetism in quark matter}

A possibility of FM in quark matter has been first suggested
by a perturbative calculation with the one-gluon-exchange (OGE)
interaction \cite{Tatsu}: the mechanism of {\it spontaneous spin
polarization} of quark
matter is caused by the Fock exchange interaction, since quark matter is
globally color neutral system, which is analogous to that due to itinerant 
electrons in condensed matter physics \cite{blo,yos}. One of the important
features there is that there is no spin independent interaction ab
initio, while the spin polarization occurs as a consequence of the Pauli
exclusion principle: electrons with the same spin direction can
effectively avoid the Coulomb repulsion. Since a quark should be treated
relativistically and no more eigenstate of the spin operator $\Sigma_z$,  
we must carefully define ``ferromagnetism'' in quark matter \cite{Tatsu}. 

By operating the Fierz transformation on the original Lagrangian, we
can extract the relevant Fock interaction to magnetism: we shall see 
the condensation of the axial-vector mean-field (AV), 
$\langle \bar{\psi} \gamma_5\gamma_3 \psi \rangle$, is a key phenomenon \cite{MaruTatsu}.
It is well known that any non-vanishing mean-field, 
$\langle \bar{\psi} \Gamma_\alpha \psi \rangle$, implies the importance
of the particle-hole (or anti-particle) degree of freedom.

Taking AV in the form,
\beq
{\bf V} = -\gamma_5 \gamma_3 {\bf U}_A , ~{\bf U}_A//\hat z ,
\label{av}
\eeq
without loss of generality, we have the quark Green function in the
presence of AV, 
\beq
G^{-1}_A(p) = \sla{p}-m-\sla{\mu}+\gamma_5 \sla{U_A}.  
\eeq
The particle spectrum is then easily obtained by seeking for the poles
of the Green function $G_A(p)$,
$\det G^{-1}_A(p_0=\epsilon_n)|_{\mu=0}=0$, for uniform magnetization:
\beqa
&& \epsilon_n=\pm \epsilon_\pm \\
  && \epsilon_{\pm}= \sqrt{{\bf p}^2+{\bf U}_A^2+m^2 \pm 2 
                        \sqrt{m^2 {\bf U}_A^2+({\bf p}\cdot {\bf U}_A)^2 }},
\eeqa
where the sign $\pm$ in $\epsilon_{\pm}$ represents an {\it exchange splitting} of 
different ``spin'' eigenstates \cite{yos}. The spectrum is reduced to a familiar form
$ \epsilon_{\pm}\sim m+\frac{p^2}{2m}\pm |{\bf U}_A|$ in the
non-relativistic limit \cite{yos}.
In the massless limit, on the other hand, $\epsilon_{\pm}\rightarrow
[p_\perp^2+(|p_z|\pm |{\bf U}_A|)^2]^{1/2}$, which is  nothing else but the
free-particle 
energy with a shift of the
$z$ component of momentum, and $U_A$ becomes a redundant degree of freedom.
We can immediately see that there appears a coupling term in $\epsilon_\pm$, 
$\propto {\bf p}\cdot {\bf U}_A$, reflecting that the direction of spin
changes as a particle moves in the relativistic case. 
\begin{wrapfigure}{r}{7.cm}
\begin{center}
\includegraphics[width=5cm,clip]{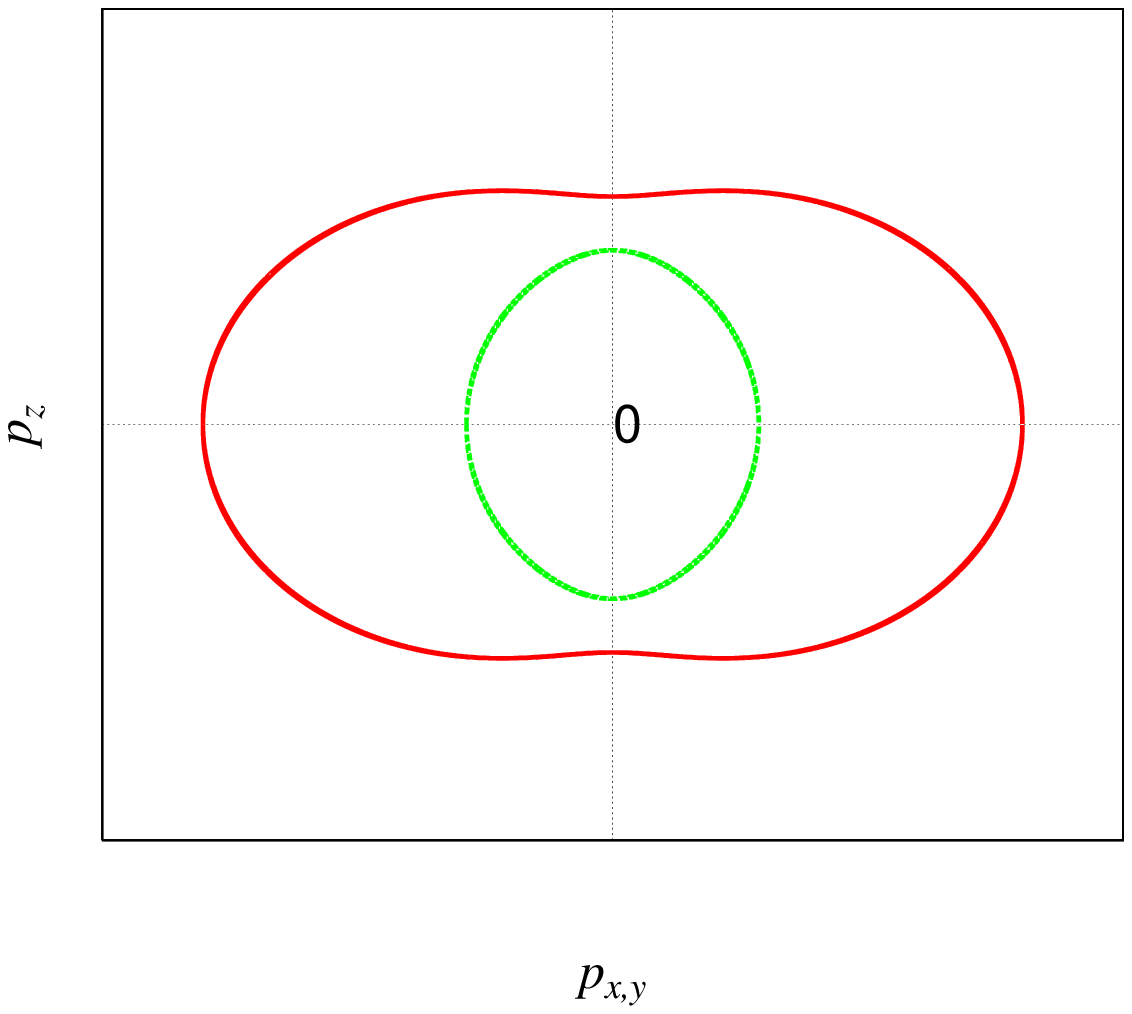}
\end{center}
\caption{Schematic view of the shapes of the Fermi seas: a prolate shape
 for majority ``spin'' quarks ($F^-$) and an oblate shape for
 minority ``spin'' quarks ($F^+$). }
\end{wrapfigure}
Thus we must consider the spin configuration in
the phase space, different from the non-relativistic case like the
Heisenberg model.

The system loses rotation symmetry in the momentum space as well as in
the coordinate space due to the presence of AV 
($ O(3)\rightarrow O(2)$), so that the Fermi sea is deformed in
the relativistic case. 
Consequently we have two deformed Fermi seas with different volumes: one ($F^-$)
has a prolate shape for majority particles and the other ($F^+$) an oblate shape
for minority particles. 
\footnote{It would be interesting to see a similar idea
given in ref.~10); these authors {\it variationally} introduced the deformed Fermi
seas in the context of the gap function
of CSC for a pair with different Fermi surfaces. 
Their shapes look very similar to our case, but its origin is
completely different to each other.}  

The mean-value of the spin operator is then given by
\beqa
{\bar s_z}=\frac{1}{2}\langle
\Sigma_z\rangle&=&-i\int_C\frac{d^4p}{(2\pi)^4}{\rm tr}
\gamma_5\gamma_3G_A(p)\nonumber\\
&=&\frac{1}{2}\left[\int_{F^+}\frac{d^3p}{(2\pi)^3} 
\frac{|{\bf U}_A|+\beta}{\epsilon_+}+\int_{F^-}\frac{d^3p}{(2\pi)^3} 
\frac{|{\bf U}_A|-\beta}{\epsilon_-}\right]
\label{magnet}
\eeqa
with $\beta=\sqrt{p_z^2+m^2}$, and we can obviously see that the
nonvanishing 
value of $|{\bf U}_A|$ causes finite magnetization.

\section{Color magnetic superconductivity}

Here let's tackle the coexistence problem of FM and CSC, a possibility of
{\it color magnetic superconductivity} \cite{nak}; more definitely,  
our main concern here is to examine the possibility of 
the quark Cooper instability under AV in Eq.~(\ref{av}), which is
responsible for spin polarization of quark matter.  In other words 
we'd like to figure out the interplay between 
particle-particle and particle-hole degrees of freedom; the former leads to
CSC, while the latter FM. 
We start with the OGE action in QCD:
\beq
    I_{int}=-g^2\frac{1}{2}\int{\rm d^4}x \int{\rm d^4}y
 \left[\bar{\psi}(x)\gamma^\mu \frac{\lambda_a}{2} \psi(x)\right]
D_{\mu \nu}(x,y)
 \left[\bar{\psi}(y)\gamma^\nu \frac{\lambda_a}{2} \psi(y)\right], 
\eeq
Using the mean-field approximation, we have 
\beq
 I_{MF}=\frac{1}{2} \int \frac{{\rm d}^4 p}{(2 \pi)^4} 
                \left( \begin{array}{l} 
                          \bar{\psi}(p)   \\
                          \bar{\psi}_c(p) \\
                       \end{array} \right)^T
                  G^{-1}(p)
                \left( \begin{array}{l} 
                          \psi(p)   \\
                          \psi_c(p) \\
                       \end{array} \right), \\
\label{mfield}
\eeq
in the Nambu-Gorkov formalism.
Assuming the color singlet particle-hole pair mean-field, $V$, as well
as the color $\bar 3$ particle-particle pair mean-field, $\Delta$, the
inverse Green function $G^{-1}$ renders 
\begin{eqnarray}
G^{-1}(p)&=&\left( \begin{array}{cc}
                      \sla{p}-m+\sla{\mu}+V(p) & 
                      \gamma_0 \Delta^\dagger(p) \gamma_0  \\
                      \Delta(p) & 
                      \sla{p}-m-\sla{\mu}+\overline{V}(p) \\
                          \end{array} \right),\nonumber\\
         &=&\left( \begin{array}{cc}
                            G_{11}(p) & G_{12}(p) \\
                            G_{21}(p) & G_{22}(p) \\
                            \end{array} \right)^{-1} \label{fullg2} 
\end{eqnarray}
where charge-conjugated spinor $\psi_c$ and mean-field $\bar V$ are
defined by 
\beq
\psi_c(k) = C \bar{\psi}^T(-k),~~~\overline{V} \equiv C V^T C^{-1}. 
\eeq
Then we immediately have the coupled equations, representing the
interplay of two different mean-fields, $V$ and $\Delta$: 
\beqa
-V(k)=(-ig)^2 \int \frac{{\rm d}^4p}{i(2\pi)^4} \{-iD^{\mu \nu}(k-p)\} 
      \gamma_\mu \frac{\lambda_\alpha}{2} \{-iG_{11}(p)\} 
      \gamma_\nu \frac{\lambda_\alpha}{2}  
\label{self1}.
\eeqa
\beq
  -\Delta(k)=(-ig)^2 \int \frac{{\rm d}^4p}{i(2 \pi)^4} \{-iD^{\mu \nu}(k-p)\}
               \gamma_\mu \frac{-(\lambda_\alpha)^T}{2}
                 \{-iG_{21}(p) \} 
               \gamma_\nu \frac{\lambda_\alpha}{2}.  
\label{gap1} 
\eeq

\begin{figure}[h]
\begin{center}
\includegraphics[width=6cm,clip]{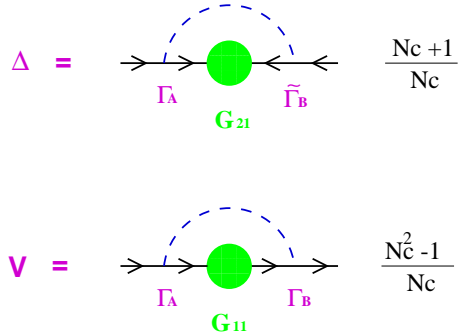}
\end{center}
\caption{Graphical interpretations of the coupled equations (3.5) and
 (3.6) with coefficients in front of R.H.S. given by $N_c$.}
\end{figure}

Applying the Fierz transformation for the self-energy term (\ref{self1})
we can see that 
there appear the color-singlet scalar, pseudoscalar, vector and axial-vector
self-energies.   
In general we must take into account these self-energies in $V$,
$V=U_s+\gamma_5U_{ps}+\gamma_\mu U_v^\mu+\gamma_\mu\gamma_5U_{av}^\mu$
with the mean-fields $U_i$. Here we retain only $U_s, U_v^0, U_{av}^3$
in $V$ and suppose that others to be vanished.

We shall see this ansatz gives 
self-consistent solutions for Eq.(\ref{self1}) because of axial and reflection
symmetries of the Fermi seas under the zero-range approximation for the
gluon propagator (see \S 3.1). We furthermore discard the scalar mean-field $U_s$ and 
the time component of the vector mean-field $U_v^0$ for simplicity 
since they are directly irrelevant for the spin degree of freedom.

Here we take the following ansatz for $V$ and $\Delta$: $V$ has a form
given in Eq.~(\ref{av}) and 
\begin{eqnarray}
\Delta({\vp})=\sum_s \tilde{\Delta}_s({\vp}) B_s({\vp}),~
B_s({\vp})=\gamma_0 \phi_{-s}({\vp}) \phi_{s}^\dagger({\vp})
\label{delta}
\end{eqnarray}
with the energy eigenspinors $\phi_s, ~s=\pm$ with eigen energies (2.4),
$\epsilon_s$, s.t. $G_A^{-1}(\epsilon_s, {\bf p})\phi_s=0$. The structure of the gap
function (\ref{delta}) is inspired by a physical consideration of a quark pair on the
same Fermi surface of $F^-$ or $F^+$ \cite{nak}: we consider here the quark pair on each Fermi surface 
with opposite momenta, ${\bf p}$ and $-{\bf p}$ so that they result in a linear combination of
$J^\pi=0^-, 1^-$ (see Fig.~3).
\footnote{Note that this choice is not unique; actually we are now
studying another possibility of quark pair between different Fermi
surfaces \cite{naw}.}
 $\tilde\Delta_s$ is still a matrix in the color-flavor space 
and we take 
\beq
\left(\tilde\Delta_s\right)_{\alpha\beta;ij}=\epsilon^{\alpha\beta 3}\epsilon^{ij}\Delta_s
\eeq
to satisfy the fermionic constraint for the two-flavor case (2SC), where
$\alpha,\beta$ 
denote the color indices and
$i,j$ the flavor indices. Then the quasi-particle spectrum can be obtained by
looking for poles of the diagonal Green function, $G_{11}$: 
\beqa
E_{s}({\vp})&&=\left\{
 \begin{array}{ll}
 \sqrt{(\epsilon_s({\vp})-\mu)^2+|\Delta_s({\vp})|^2} & \mbox{for color 1, 2} \\
 \sqrt{(\epsilon_s({\vp})-\mu)^2}            & \mbox{for color 3} 
 \end{array} 
        \right.  
\label{qusiE}
\eeqa
Note that the quasi-particle energy is independent of color and flavor
in this case, since we have assumed a singlet pair in flavor and color. 
Gathering all these stuffs to put them in the self-consistent
equations (3.5), (3.6), we definitely write down the coupled equations for $\Delta_s$,
\beqa
 \Delta_{s'}(k,\theta_k)=\frac{N_c+1}{2N_c} \tilde{g}^2 
 \int \frac{{\rm d}p\, {\rm d}\theta_p}{(2\pi)^2} p^2 \sin\theta_p
   \sum_s T_{s' s}(k,\theta_k,p,\theta_p) 
   \frac{\Delta_s(p,\theta_p)}{2 E_s(p,\theta_p)}, \label{GAP1}  
\eeqa
and for $U_A$,
\beqa
  U_A
  &=& -\frac{N_c^2-1}{4N_c^2}\tilde{g}^2 \int \frac{{\rm d}^3 p}{(2\pi)^3} 
\sum_s \left\{ \theta(\mu-\epsilon_s(\vp)) + 2 v_s^2({\vp})\right\}
\frac{U_A +s \beta_p}{\epsilon_s(\vp)}, 
\label{UA1} 
\eeqa
within the ``contact'' interaction with $\tilde g^2=g^2/\Lambda^2$ (see Eq.~(\ref{contact})), 
where $v_s^2(\vp)$ denotes the momentum distribution of the
   quasi-particles (cf. Eq.~(\ref{magnet})).
Carefully analyzing the structure of the function $T_{s's}$ in
   Eq.~(\ref{GAP1}),
 we can easily find that the gap function s.t. 
   $\Delta_s$ should have the polar angle ($\theta$) dependence on the Fermi
   surface s.t. 
\beq
\Delta_s(p^F_s,\theta)=\frac{p^F_s(\theta)\sin \theta}{\mu}
\left( -s\frac{m}{\sqrt{m^2+( p^F_s(\theta) \cos \theta )^2}} R + F \right),
\label{pola}
\eeq
with constants $F$ and $R$ to be determined.
\begin{figure}[h]
\begin{center}
\includegraphics[width=5cm,clip]{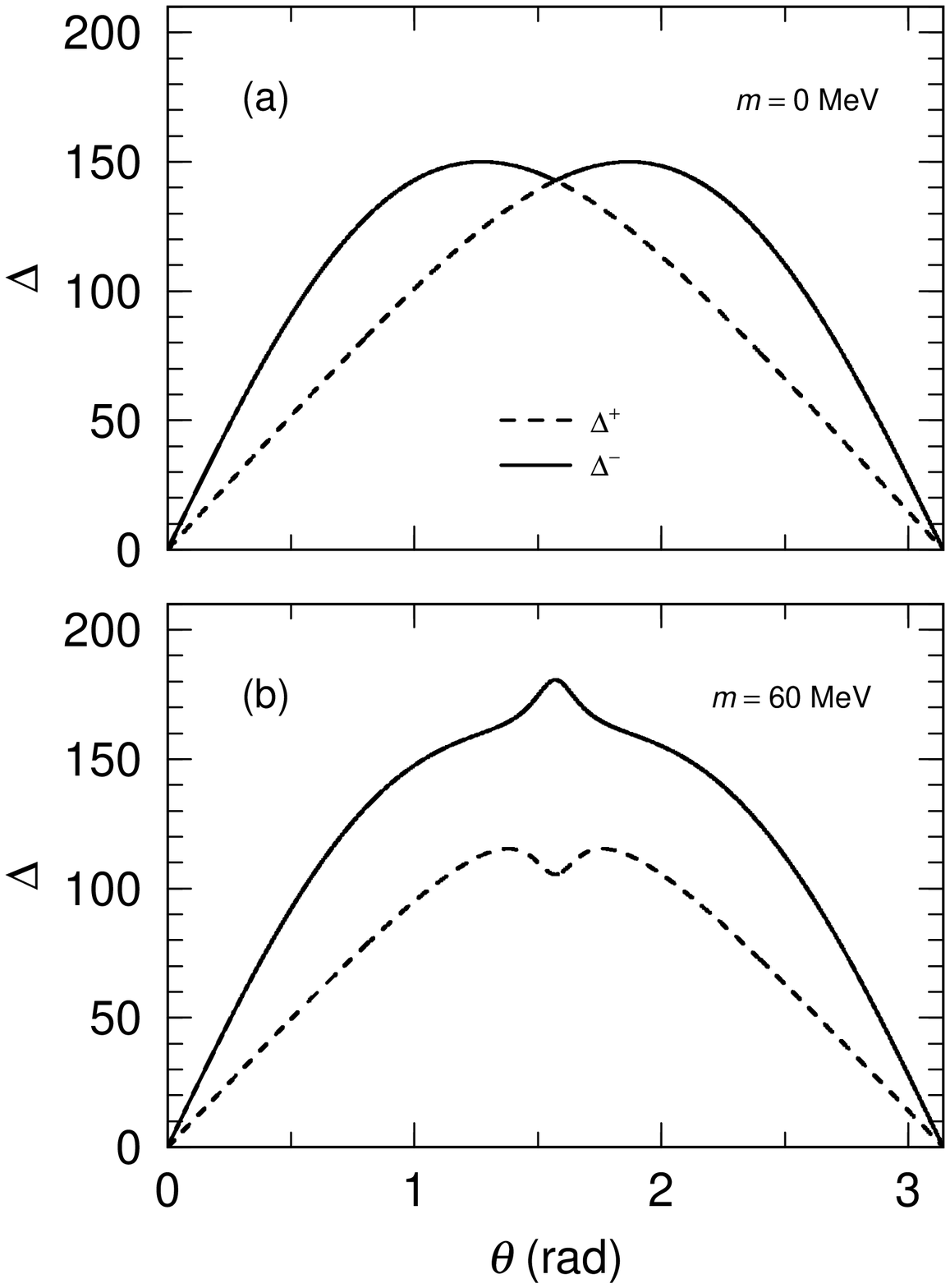}
~~~~~~~~~~~~\includegraphics[width=6cm,clip]{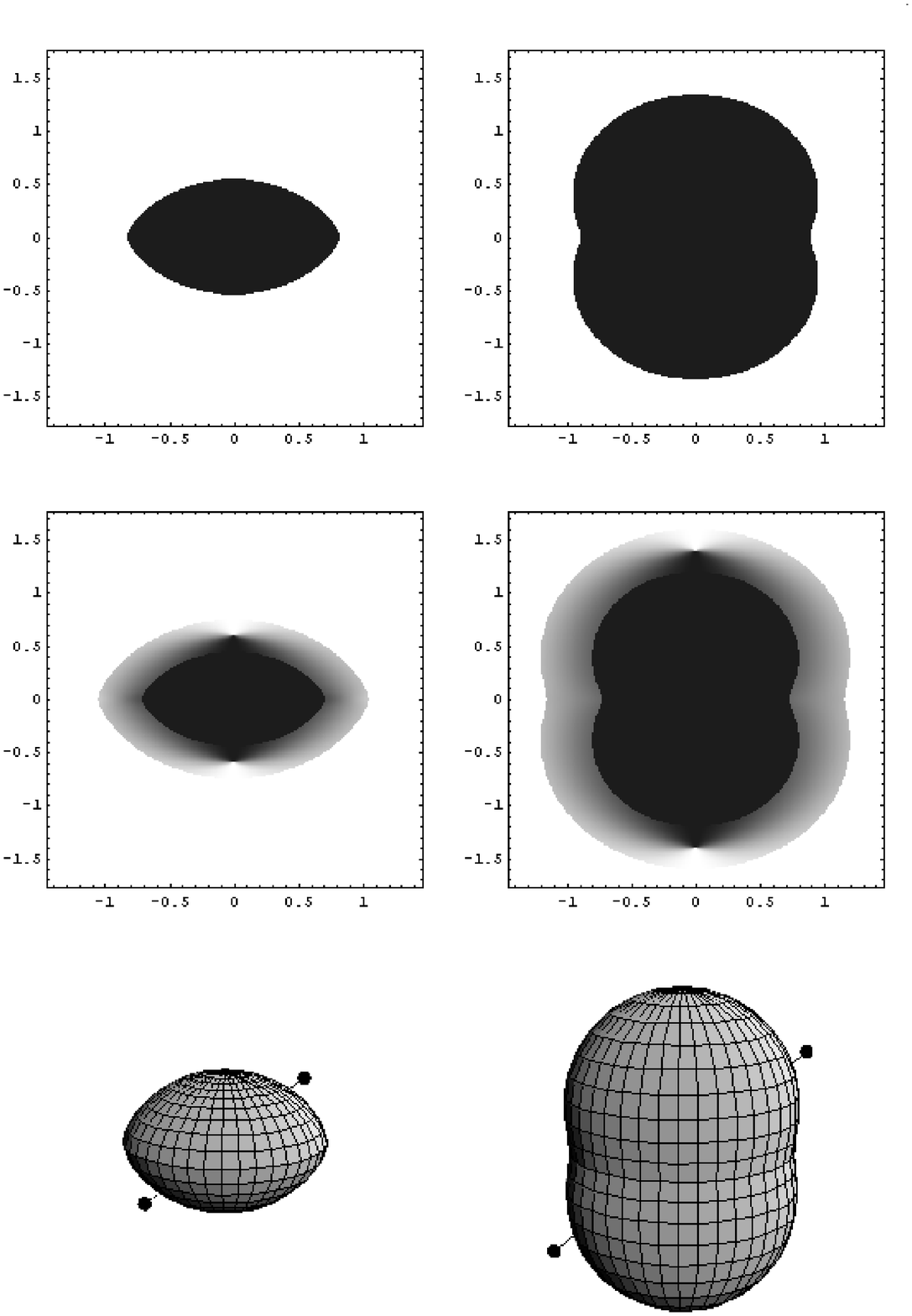}
\end{center}
\caption{ (left panel) Schematic view of the polar-angle dependence of the gap
 functions at the Fermi surface: (a) for $m=0$ and (b) for $m\neq 0$.
(right panel) Deformed Fermi seas and quark pair on each surface. The middle
 two figures show diffusion of the Fermi surfaces in the presence of 
 the polar-angle dependent gap function $\Delta_\pm$.}
\end{figure}
As a characteristic feature, both the gap functions have nodes at poles
($\theta=0,\pi$) and take the maximal values at the vicinity of equator
($\theta=\pi/2$), keeping the relation, $\Delta_-\gtrsim \Delta_+
$ (Fig.~3). This feature is very similar to the $^3 P$ pairing in liquid $^3$He or
nuclear matter \cite{He3,NM3P}; actually we can see our gap function 
Eq.~(\ref{pola}) to exhibit an 
effective $P$ wave nature by a genuine relativistic effect due to the Dirac spinors.
To summarize, we depict in Fig.~(3) the deformed Fermi seas for
$\Delta_\pm=0$ and the quasi-particle
distributions for $\Delta_\pm\neq 0$.

We also find that the expression for $U_A$,
   Eq.~(\ref{UA1}), is nothing but the simple sum of the expectation
   value of the spin operator with the weight of the occupation
   probability of the quasi-particles $v_s^2$ for two colors and the
   step function for remaining one color (cf. (\ref{magnet})). 

\subsection{Self-consistent solutions}

Here we demonstrate some numerical results; we replaced the original OGE
by the ``contact'' interaction with the cutoff around the Fermi surface
in the momentum space,
\begin{eqnarray}
D^{\mu\nu}\rightarrow -g^{\mu\nu}/\Lambda^2,~
\Delta_s({\bf p}) 
\rightarrow \Delta_s({\bf p})\theta(\delta-|\epsilon_s-\mu|)
\label{contact}
\end{eqnarray}
as in the BCS theory in the weak-coupling limit.




First we show the magnitude of AV.
It is seen that the axial-vector mean-field (spin polarization) 
appears above a critical density and
becomes larger as baryon number density gets higher.
Moreover, 
the results for different values of the quark mass show that
spin polarization grows more for the larger quark mass.
This is because a large quark mass gives rise to much difference 
in the Fermi seas of two different ``spin'' states, 
which leads to growth of the exchange energy in the axial-vector channel.
A slight reduction of $U_A$ arises as a result of diffuseness of the
Fermi surface due to $\Delta_s$. As seen in Eq.~(\ref{UA1}), $U_A$ can
be obtained as addition and cancellation of the contributions by two different Fermi seas;
the latter term is more momentum dependent than the former one and
thereby $v_s^2({\bf p})$ enhances the cancellation term (see Fig.~4).

\begin{figure}[h]
\begin{center}
 \includegraphics[width=5cm,clip]{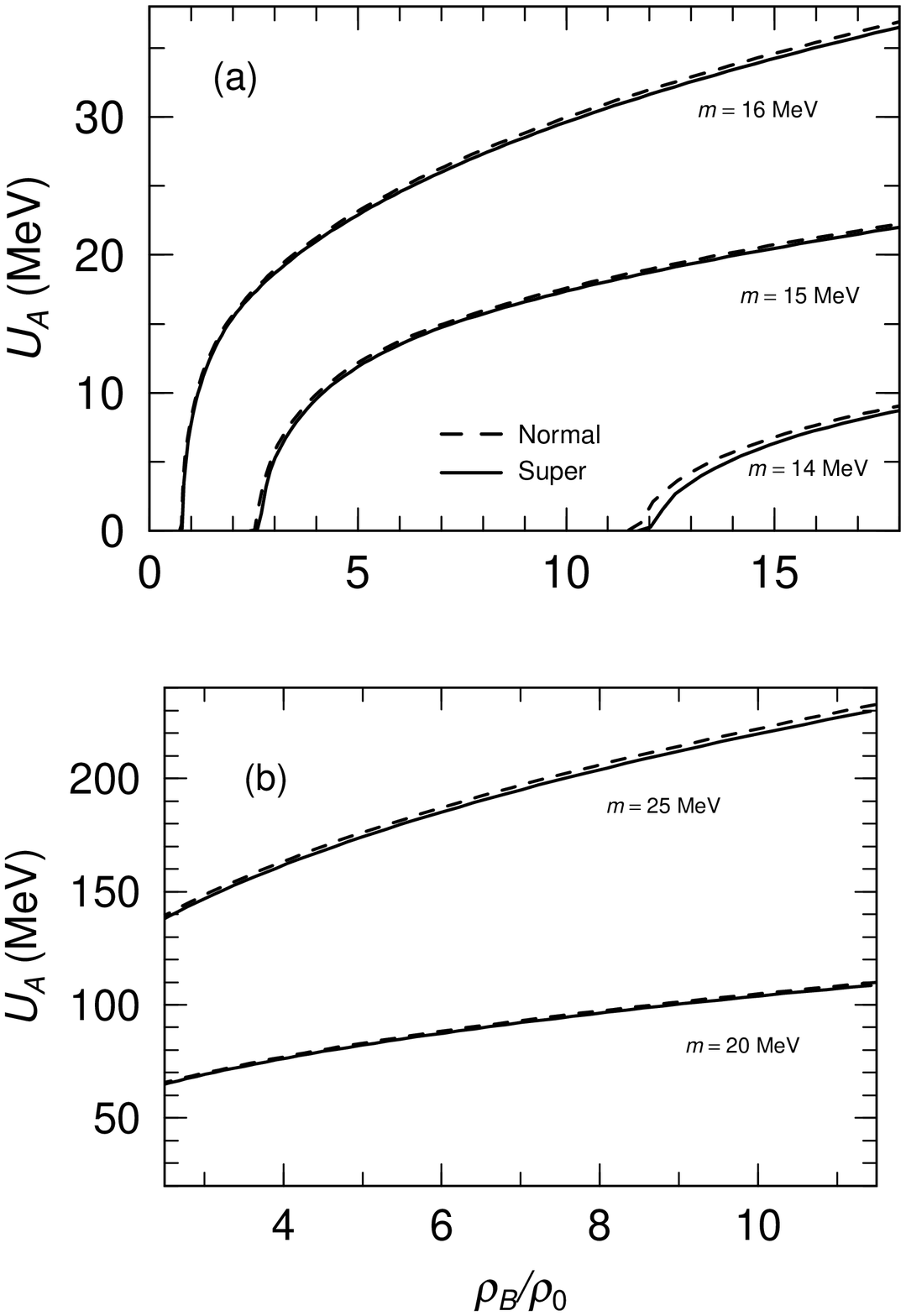}
 \includegraphics[width=5cm,clip]{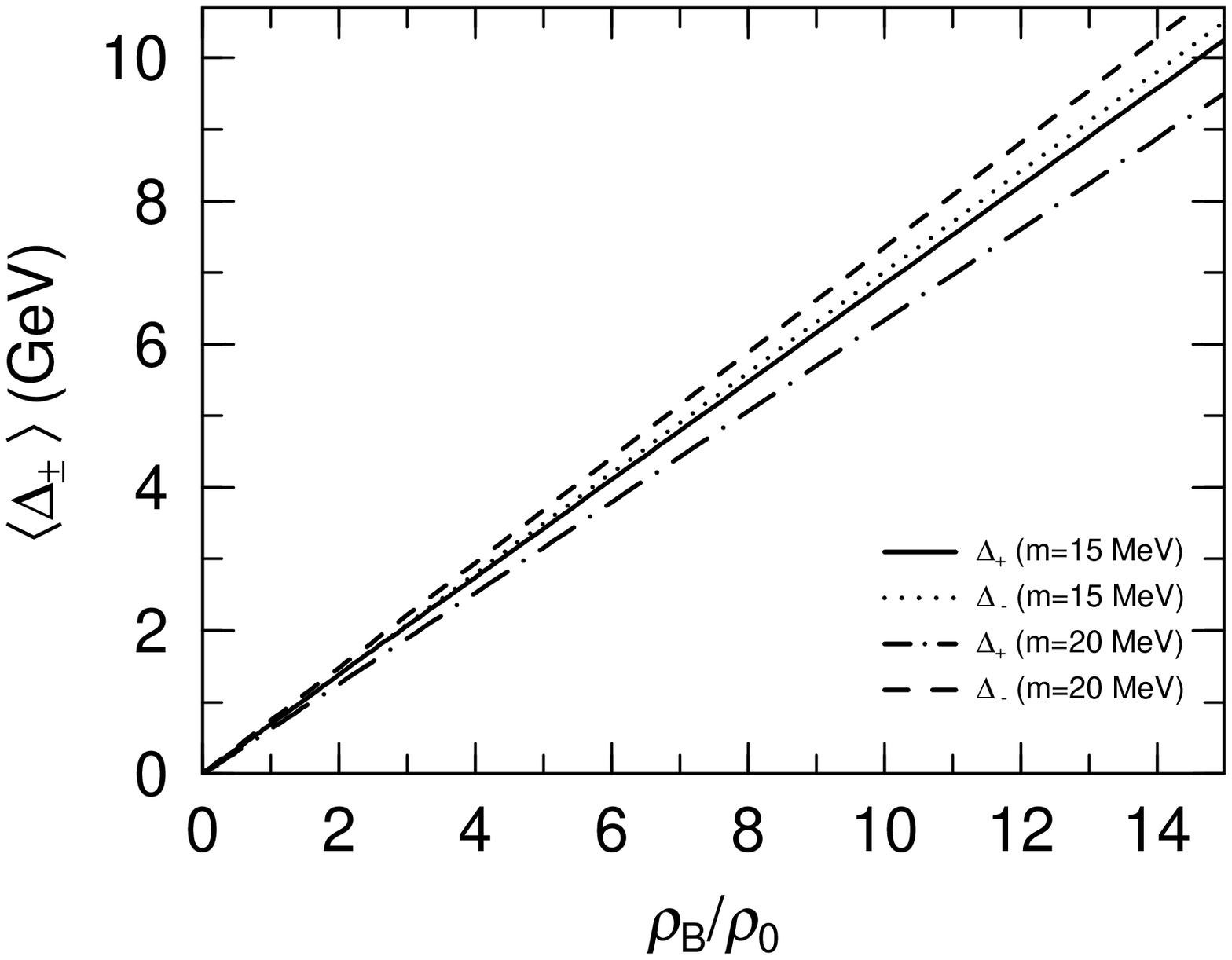}
\caption{(Left panel) Axial-vector mean-field (AV) as a function of baryon number density
 $\rho_B$($\rho_0=0.16$fm$^{-3}$). Solid (dashed) lines denote AV in the
 presence (absence) of CSC; (Right panel) Mean value of the gap functions.}
\end{center}
\end{figure}

Next we show the gap function as a function of $\rho_B$. To see the bulk
behavior of the gap function, we use the mean-value with respect to the
polar angle on the Fermi surface,
$
\langle \Delta_\pm \rangle \equiv 
\left( \int_0^\pi {\rm d}\theta \sin\theta\Delta_\pm^2/2 \right)^{1/2}.
$
The mean values $\langle \Delta_\pm \rangle$ begin to split at a 
certain density where $U_A$ becomes finite.

With these figures we can say that FM and CSC barely interfere with each
other.

%

\section{Summary and concluding remarks}

We have discussed color magnetic superconductivity with a simple model
and found that the coexistence of FM and CSC is possible, while CSC
somewhat suppresses FM. 

We can still expect a high magnetic field enough
to explain $B_{\rm MAX}\sim 10^{15}$G observed in magnetars, while 
the magnitude of polarization is not so large ($\sim $ several $\%$).

Through numerical calculations we used the constant quark masses  
by regarding them as input
parameters and studied the mass effect on the present problem. We found that 
the quark mass is very important to realize, especially, FM, and  a
larger quark mass favors spontaneous magnetization.
Once taking into account the restoration of chiral symmetry at a certain 
density, we must allow a dynamical change of the quark mass. This
subject is under progress \cite{tatnak}.  

Finally it would be worth mentioning related
other works. As a recent progress in CSC, some authors considered 
$S=1$ pairing as well as a usual one \cite{bub}, 
which also exhibits an polar-angle dependence of the gap
function. However, its magnitude is much less than the usual one. Note
that there is only one type of pairing in our case.

We hope this work serves as an impetus for further study on color
magnetic superconductivity.

\section*{Acknowledgements}

The present work of T.T. and T.M. is partially supported by the 
Japanese Grant-in-Aid for Scientific Research Fund of the Ministry
of Education, Culture, Sports, Science and Technology (11640272,
13640282), and by the REIMEI Research
Resources of Japan Atomic Energy Research Institute (JAERI).

\end{document}